\documentclass[aps,prl,twocolumn,showpacs,superscriptaddress]{revtex4}
\usepackage{appendix} 
\usepackage{graphicx}
\usepackage{amsmath}
\usepackage{amssymb}
\usepackage{colordvi}
\label{default}
\usepackage{mathrsfs}
\usepackage{bm}
\usepackage{verbatim}
\usepackage{dcolumn}
\usepackage{bm}
\usepackage{epsfig}
\usepackage{subfigure}
\begin{document}
\title{Single-Valley Engineering in Graphene Superlattices}
\author{Yafei Ren}
\affiliation{International Centre for Quantum Design of Functional Materials, Hefei National Laboratory for Physical Sciences at Microscale, and Synergetic Innovation Center of Quantum Information and Quantum Physics, University of Science and Technology of China, Hefei, Anhui 230026, China}
\affiliation{Department of Physics, University of Science and Technology of China, Hefei, Anhui 230026, China}
\author{Xinzhou Deng}
\affiliation{International Centre for Quantum Design of Functional Materials, Hefei National Laboratory for Physical Sciences at Microscale, and Synergetic Innovation Center of Quantum Information and Quantum Physics, University of Science and Technology of China, Hefei, Anhui 230026, China}
\affiliation{Department of Physics, University of Science and Technology of China, Hefei, Anhui 230026, China}
\author{Changsheng Li}
\affiliation{Department of Physics, Hunan University of Arts and Science, Changde, Hunan 415000, China}
\author{Jeil Jung}
\affiliation{Graphene Research Centre and Department of Physics, National University of Singapore, 2 Science Drive 3, Singapore 117551, Singapore}
\author{Changgan Zeng}
\affiliation{International Centre for Quantum Design of Functional Materials, Hefei National Laboratory for Physical Sciences at Microscale, and Synergetic Innovation Center of Quantum Information and Quantum Physics, University of Science and Technology of China, Hefei, Anhui 230026, China}
\affiliation{Department of Physics, University of Science and Technology of China, Hefei, Anhui 230026, China}
\author{Zhenyu Zhang}
\affiliation{International Centre for Quantum Design of Functional Materials, Hefei National Laboratory for Physical Sciences at Microscale, and Synergetic Innovation Center of Quantum Information and Quantum Physics, University of Science and Technology of China, Hefei, Anhui 230026, China}
\author{Qian Niu}
\affiliation{Department of Physics, The University of Texas at Austin, Austin, Texas 78712, USA}
\affiliation{International Center for Quantum Materials and Collaborative Innovation
Center of Quantum Matter, Peking University, Beijing 100871, China}
\author{Zhenhua Qiao}
\email[Correspondence author:~~]{qiao@ustc.edu.cn}
\affiliation{International Centre for Quantum Design of Functional Materials, Hefei National Laboratory for Physical Sciences at Microscale, and Synergetic Innovation Center of Quantum Information and Quantum Physics, University of Science and Technology of China, Hefei, Anhui 230026, China}
\affiliation{Department of Physics, University of Science and Technology of China, Hefei, Anhui 230026, China}
\date{\today}

\begin{abstract}
  The two inequivalent valleys in graphene preclude the protection against inter-valley scattering offered by an odd-number of Dirac cones characteristic of $\mathbb{Z}_2$ topological insulator phases. Here we propose a way to engineer a chiral single-valley metallic phase with quadratic crossover in a honeycomb lattice through tailored  $\sqrt{3}N\times\sqrt{3}N$ or $3N\times 3N$ superlattices. The possibility of tuning valley-polarization via pseudo-Zeeman field and the emergence of Dresselhaus-type valley-orbit coupling are proposed in adatom decorated graphene superlattices. Such valley manipulation mechanisms  and metallic phase can also find applications in honeycomb photonic crystals.
\end{abstract}
\maketitle

\textit{Introduction---.}
Honeycomb Dirac materials have two-fold degenerate band structures with inequivalent KK$'$ valleys~\cite{graphene1,graphene2,spintronics,spin-shen,graphene-spin}, whose origin can be traced back to the bipartite nature of honeycomb lattices (A and B triangular sublattices). This binary valley degree of freedom has led to proposals of valleytronics applications~\cite{Kane,Qiao1,Rycerz,Guinea,JianLi,Pesin}
that leverage the valley pseudospins in a manner analogous to electron spins in spintronics applications. A distinct scenario is that of single (odd-number) Dirac-cone in $\mathbb{Z}_2$ topological insulators \cite{hasanTI} where their surface states are effectively decoupled from each other due to their distant spatial separation. Therefore, a single Dirac-cone structure is desirable when we require a Hamiltonian that embodies the chiral anomaly of Dirac fermions \cite{semenoffchiral} and at the same time is protected against inter-valley scattering.

In this Letter, we propose to engineer a single valley phase in 2D honeycomb Dirac materials through $\sqrt{3}N\times\sqrt{3}N$ or $3N\times 3N$ superlattices that fold and couple the inequivalent KK$'$ valleys into the same $\Gamma$ point. We show that the corresponding effective Hamiltonians for top-site adsorbed superlattices exhibit uniform inter-valley coupling and valley-orbit coupling mechanisms that resemble the conventional in-plane Zeeman fields and Dresselhaus spin-orbit coupling of the electron spins~\cite{spintronics,spin-shen,spinFET,Qiao2,Qiao3,WuRuqian1,WuRuqian2,ZhangHongbin}. The pseudo-Zeeman field and pseudospin-orbit coupling allow to control valley polarization coherently, while the latter one further indicates the possibility of controlling valley polarization via electric fields. Moreover, together with the coexisting sublattice potentials, we find that inter-valley coupling can drive a topological phase transition from a quantum valley-Hall phase into a chiral single-valley metallic phase with quadratic band crossover. We also propose that such inter-valley coupling mechanism and metallic phase can be observed in photonic crystals.

\begin{figure}
  \includegraphics[width=8.5cm,angle=0]{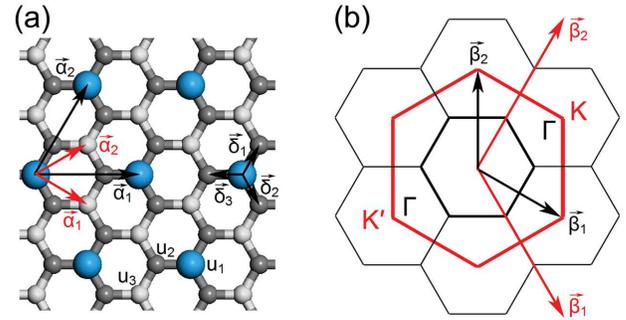}
  \caption{  \label{figure1} (color online)
  Schematic representation of inter-valley coupling adatom superlattices and their respective Brillouin zones.
  (a) and (b) are respectively primitive and reciprocal lattices for the top adsorption in $\sqrt{3}\times\sqrt{3}$ graphene supercells. The red lines represent the Brillouin zone of pristine graphene.}
\end{figure}

\textit{Inter-Valley Coupling---.}
When the $\sqrt{3} N\times \sqrt{3}N$ or $3 N\times 3 N$ supercells are tailored on a honeycomb lattice, KK$'$ valleys couple and fold into the $\Gamma$ point,
as illustrated schematically in Fig.~\ref{figure1}(b) showing the reciprocal lattices for both $1\times 1$ (red) and $\sqrt{3} \times \sqrt{3}$ (black) supercells. For definiteness, here we only focus on the top-site adsorption as shown in Fig.~\ref{figure1}(a) and leave the discussion of the effective Hamiltonians for bridge- and hollow-site adsorption in the Supplemental Material. For top-site adsorption in a $\sqrt{3} \times \sqrt{3}$ supercell, the six atoms in each primitive cell can be classified into three different categories: (i) one at the adatom site, (ii) three at the nearest neighbor sites,  and (iii) two at next-nearest sites. We represent the corresponding site energies as $u_1$, $u_2$, and $u_3$, and set $u_3=0$ as the reference value. Assuming that the adsorption sites belong to sublattice ``A", the real-space tight-binding Hamiltonian can be written as:
\begin{eqnarray}
           H_{\rm{t}} = H_0  +  u_1 {\sum_{i}}' a_{i}^{\dag} a_{i} + u_2 {\sum_{i}}'\sum_{\delta} b_{i+\delta}^{\dag} b_{i+\delta},
           \label{eq1}
\end{eqnarray}
where ${\sum_{i}}'$ runs over all adatom sites. Here $H_0=- t_0 \sum_{<ij>}(a_i^{\dag}b_j + h.c.)$ is the band Hamiltonian with $t_0$ being the nearest-neighbor hopping energy, and $a_i^{\dag}$ ($b_i^{\dag}$) is the creation operator of an electron at $i$-th A(B) site.

The Brillouin zone of pristine graphene can be represented through three copies of $\sqrt{3}\times \sqrt{3}$ graphene supercell's Brillouin zone as displayed in Fig.~\ref{figure1}(b). By denoting $j$-th ($j$=1-3) center as ${\bf{K}}_j$, the operator $a_i$ can be expanded in momentum space as:
$a_i = \frac{1}{\sqrt{N_0}} \sum_{\bm k} \sum_j \exp [{-i ({\bf{K}}_j+{\bm k}) \cdot {\bf{R}}_i }]a_{j,k}$, where $N_0$ is a normalization factor, and $\bm{k}$ runs over the Brillouin zone of $\sqrt{3}\times \sqrt{3}$ graphene supercell. The Hamiltonian of Eq.~(\ref{eq1}) in momentum space is:
\begin{eqnarray}\label{Expansion-k-s3xs3}
           H_{\rm{t}}(\bm{k}) =H_0(\bm{k})+ \sum_{j,j'} [\frac{u_1}{3} a_{j,k}^{\dag}a_{j',k} + \frac{u_2}{3} \xi_{jj'} b_{j',k}^{\dag}b_{j,k}],
\end{eqnarray}
where $H_0(\bm{k})=- t_0 \sum_{j}(\chi_{jk} a_{j,k}^{\dag}b_{j,k} + h.c.)$ describes the kinetic energy of pristine graphene with $\chi_{jk}=\sum_{\delta}e^{-i({\bf{K}}_j+\bm{k})\cdot \bm{\delta}}$, and $\xi_{jj'}=\sum_{\delta}e^{i({\bf{K}}_j-{\bf{K}}_{j'})\cdot \bm{\delta}}$.
The ${\bf{K}}_j$ ($j$=1-3) are respectively wavevectors of K, K$'$, and $\Gamma$ points. The last two terms give sublattice potentials when $j=j'$ which are different for AB sublattices due to inversion symmetry breaking. When $j\neq j'$, they give rise to inter-valley coupling through a finite $u_1$ contribution while $u_2$ contribution vanishes due to the phase interference ($\xi_{KK'}$=0). By block diagonalization, the low-energy effective Hamiltonian can be further obtained:
\begin{eqnarray}
           H^{\rm eff}_{\rm{t}}&=&U_0+ v_F(k_x \sigma_x + \tau_z k_y \sigma_y) + \Delta_1 \sigma_z \\ \nonumber
            &+& \frac{\Delta_2}{2} (1+\sigma_z) \tau_x, \label{eff}
\end{eqnarray}
where $U_0$=$(\Delta_2+u_2)/2$ and $\Delta_1$=$(\Delta_2-u_2)/2$ with $\Delta_2$= $u_1/3$. The third term reflects the effective potential imbalance through a mass term of magnitude $\Delta_1$ and the last term describes inter-valley coupling through the $\tau_x$ operator. We note that the coupling between K and K$'$ valleys only occurs at ``A" sublattice with the coupling amplitude $\Delta_2$ depending on $u_1$ linearly. Such an inter-valley coupling acts on the valley pseudospin as an effective Zeeman field that can be used to control the valley polarization coherently in valleytronics devices.

When the nearest neighbor hopping terms of superlattice Hamiltonians are allowed to change by $\delta t=t-t_0$ due to the influence of the adatoms, the real-space tight-binding Hamiltonian in Eq.~(\ref{eq1}) acquires an additional term $H'={\sum_{\langle i,j\rangle}^{\prime}}{\delta t (a^{\dag}_i b_j}+h.c.)$
where the index $i$ runs over ``A" sites right underneath the adatoms and the $j$ sites represent the three nearest ``B" sites~\cite{SeeSI}. The modified effective Hamiltonian becomes:
\begin{eqnarray}
    {H_{\rm{t}}^{\rm eff}}'({\bm k })&=&{U_0'}+v'_F(\sigma_x {\bm 1}_\tau k_x + \sigma_y \tau_z k_y)+\Delta_1' \sigma_z {\bm 1}_\tau \nonumber \\
&+& \frac{\Delta^{\prime}_2}{2} ({\bm 1}_\sigma+\sigma_z) \tau_x + v_\delta \sigma_x(\tau_x k_x-\tau_y k_y), \label{topeff-1}
\end{eqnarray} where ($U_0'$, $\Delta_1'$) have same forms as ($U_0$, $\Delta_1$) by changing $\Delta_2$
to be $\Delta_2'=3u_1 t_0^2/(t+2t_0)^2$, and the Fermi velocity is modified to be $v'_F=v_F(2t+t_0)/(t+2t_0)$. The last term in Eq. (\ref{topeff-1}) can be identified as a Dresselhaus-type valley-orbit interaction of strength $v_\delta=v_F(t-t_0)/(t+2t_0)$ coupled with a sublattice-flip potential. This term also couples different valleys and implies the possibility of manipulating the valley degree of freedom by external electric field in a manner analogous to the control of electron spin by electrical means via spin-orbit coupling.

\begin{figure}
  \includegraphics[width=8.5cm,angle=0]{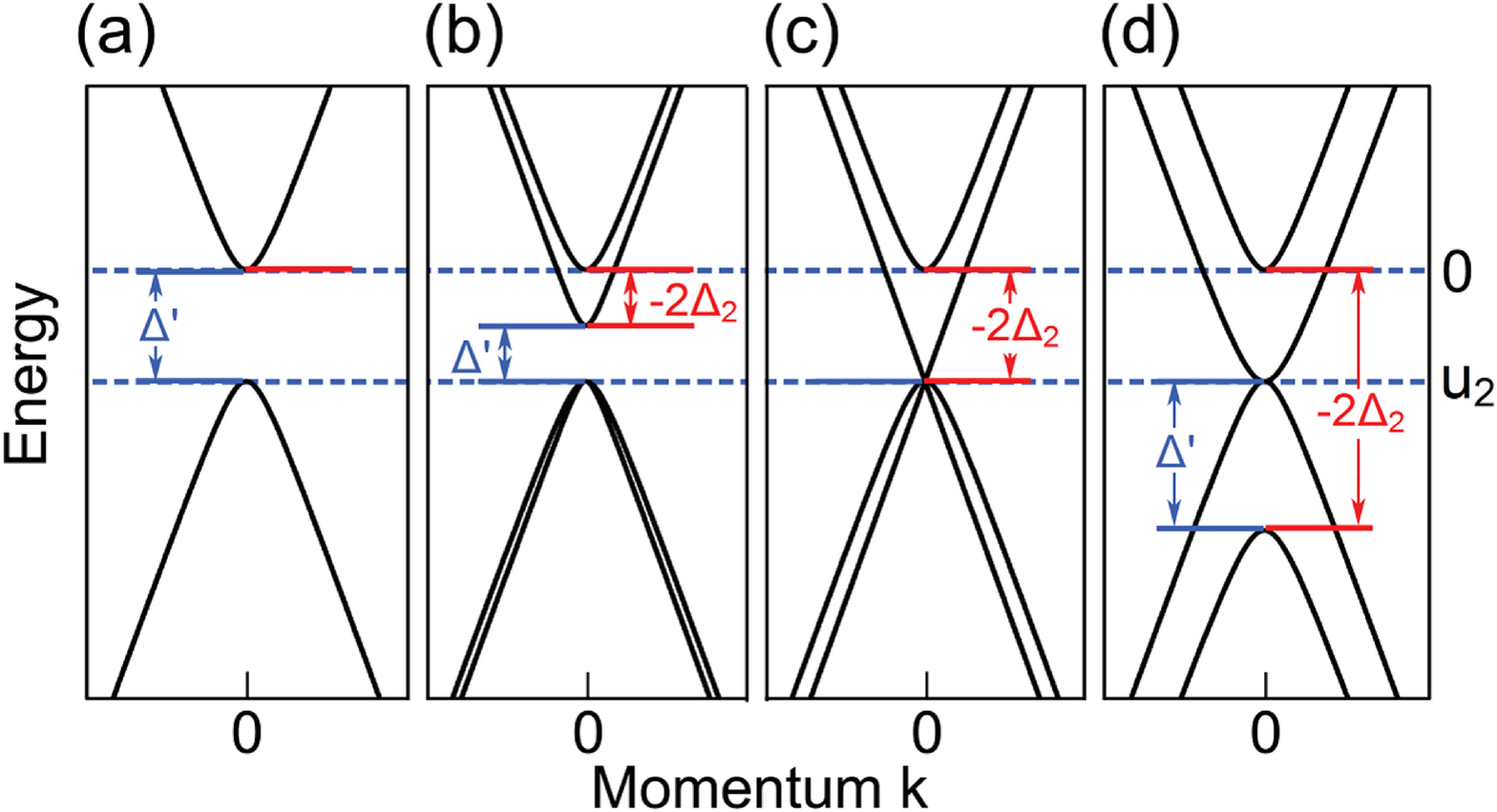}
  \caption{(color online)
  Topological transition from a quantum valley-Hall insulator to a single-valley phase as a function of the parameters  $u_1$. Here, we set $u_2$ to be fixed with $u_2<0$ and $u_3=0$.  $-2\Delta_2$ corresponds to the local band gap from the inter-valley scattering. $\Delta'$ measures the bulk (local) band gap from the competition between inter-valley coupling and sublattice potentials. The progressive decrease of $u_1$ leads to a complete closure of the quantum valley-Hall gap and then transitions to the single valley phase by reversing $\Delta'$.
  }\label{figure2}
\end{figure}

\textit{Single-Valley Metallic Phase---.} Adatom superlattices lead to both inter-valley coupling and inversion symmetry breaking potentials, and it is easy to understand that each term can independently contribute in opening a Dirac point gap when they are viewed as uniform in-plane $xy$ and $z$ contributions to the pseudospin fields in the Dirac Hamiltonian~\cite{gbn}, where the former shifts the position of the Dirac points in momentum space and the latter introduces an inversion symmetry breaking gap in the Dirac cone. Here we show that when those effects are present in a superlattice, a topologically distinct single-valley phase can be engineered. We begin considering for sake of clarity the top-adsorption configurations neglecting the modification of the hopping energy in the band Hamiltonian and setting the site energies at all ``B" sublattices to assume a constant value (\textit{i.e.}, $U_B=u_2<0$). When $u_1=0$, the site energies at all ``A" sublattices are identical, \textit{i.e.}, $U_A=0$. This leads to vanishing inter-valley scattering and the imbalanced sublattice potentials open a quantum valley-Hall gap at the Dirac points, where the doubly-degenerate massive Dirac cones are folded as a single valley around the $\Gamma$ point but remain distinguishable [See Fig.~\ref{figure2}(a)]. When we allow $u_1$ to take negative values, we find a gradual decrease of the inversion symmetry breaking induced gap $|\Delta_1|$ and an increase of inter-valley coupling strength $|\Delta_2|$ that lifts the degeneracy of the conduction bands splitting by a magnitude of $2\Delta_2$ [See Fig.~\ref{figure2}(b)]. The simultaneous presence of both terms breaks the particle-hole symmetry and leads to a smaller bulk gap
$\Delta'=|2\Delta_1+\Delta_2|$.

When $u_1$ is even further decreased and reaches a critical value of $u_1=3u_2/2$, the bulk gap $\Delta'$ completely closes. As shown in Fig.~\ref{figure2}(c), we achieve a single band touching point at $\Gamma$ formed by a Dirac-cone centered at the edge of the parabolic valence band. In this limit where the bulk gap is closed, the valley-Hall effect is absent and the valleys are no longer distinguishable. When we allow even smaller values of $u_1$, the inter-valley coupling strength $|\Delta_2|$ further increases, while the magnitude of the staggered sublattice potentials $|\Delta_1|$ first decreases to zero then increases again~\cite{SeeSI}. When the inter-valley coupling is strong enough, a valley-mixed metallic phase with quadratic band crossover is engineered as displayed in Fig.~\ref{figure2}(d). In this limit the edges of the lower energy bands are distant from the corssing point by $\Delta^\prime = \Big{|}2|\Delta_1|-|\Delta_2|\Big{|}$ [see Supplemental Material for details of the tight-binding band structure].

A detaild analysis of the low energy bands reveals that the parabolic dispersion at $\Gamma$ point is chiral and formally similar to the band dispersion near K/K$'$ valley in Bernal-stacking bilayer graphene. When $|\Delta_2| \gg |\Delta_1|$, the low-energy Hamiltonian of the quadratic touching bands can be further simplified as:
\begin{eqnarray}
    {H_{\rm{t}}^{\rm eff}}''({\bm k })&=&{U_0''}+\alpha k^2 - \beta
    \left[\begin{array}{cc}
     0     & (\pi^{\dag})^2 \\
     \pi^2 & 0
    \end{array}
     \right], \label{topeff-2}
\end{eqnarray}
which is represented on the basis of ``B" sublattice from both K and K$'$ valleys. Here, we define $U_0''=U_0-\Delta_1$, $k^2=k_x^2+k_y^2$, $\alpha=\Delta_1 v_F^2/(\Delta_2^2-\Delta_1^2)$, and $\beta=\Delta_2 v_F^2/(\Delta_2^2-\Delta_1^2)$. The last term couples states between valleys K and K$'$ with $\pi=k_x+ik_y$, and gives rise to the quadratically dispersing Fermi point band structure. Such a two-fold degeneracy at the crossing point, which also appears in Figs.~\ref{figure2}(a)-\ref{figure2}(c), is protected by the $C_{3v}$ symmetry, since these two basis functions form a two-dimensional irreducible representation of the corresponding point group. For other adsorption geometries, \textit{e.g.}, hollow- and bridge-adsorption with respectively $C_{6v}$ and $C_{2v}$ symmetries, the band structures as well as inter-valley coupling mechanisms become completely different as listed in Table~\ref{tab1}~\cite{SeeSI}.

\begin{table}
  \caption{Inter-valley coupling mechanisms for different adsorption geometries.}
  \begin{tabular}{c|c|c}
  \hline
  \hline
  Adsorption Site & Symmetry & Inter-valley Coupling      \\ \hline
  Top             & $C_{3v}$ & $(1+\sigma_z) \tau_x    $  \\ \hline
  Hollow          & $C_{6v}$ & $\tau_x\sigma_y         $  \\ \hline
  Bridge          & $C_{2v}$ & $\tau_x{\bm{1}}_{\sigma}$  \\ \hline
  \hline
  \end{tabular}
  \label{tab1}
\end{table}

The main difference of this band crossover from the case of bilayer graphene is that here we have only a \textit{single} Dirac parabolic dispersion. This is of interest, because it provides an ideal platform to study the single Dirac-cone transport phenomena of $\mathbb{Z}_2$ topological insulators and allows to explore the chiral anomaly of single valley physics that is not compensated by its time-reversal counterpart. For example, if broken symmetry gapped phases are developed in the presence of electron-electron interactions~\cite{abrikosov,hongki,jeil}, a mass sign dependent spontaneous orbital moments will develop per spin-valley~\cite{xiaodi,FanZhang,jeil}. In our single valley phase, it is expected that when the Fermi surface lies at the crossing point, a quantum anomalous Hall ground state will develop when both spin components have the same mass, or alternatively a quantum spin-Hall state will be present when the masses for each spin term have opposite signs~\cite{Murray}. Besides, a superconducting phase can also be expected when the Fermi surface is shifted away from the crossing point~\cite{Pawlak}. Whereas the energetically favored ground state depends on details of the band Hamiltonian and the models for the electron-electron interaction, further control of quantum phase transitions should be achievable by means of external magnetic fields coupling with the spontaneous orbital moments. Furthermore, in bilayer graphene, the magnitude of the gaps predicted in a Hartree-Fock theory without dynamical screening is on the order of a few tens of meV~\cite{jeil} whereas experimental gaps turned out to be an order of magnitude smaller $\sim$2~meV~\cite{jairo} due to the exponentially increasing screening feedback when the gaps are small. Thus, it is expected that substantially larger gaps can develop, if flatter bands can be tailored when the leading parabolic dispersion coefficients can be made smaller than the one used in bilayer graphene. Moreover, in the presence of strong magnetic field, the anomalous Landau-level quantization can also be expected as that in bilayer graphene case~\cite{AQHE}.

\begin{figure}
  \includegraphics[width=8.5cm,angle=0]{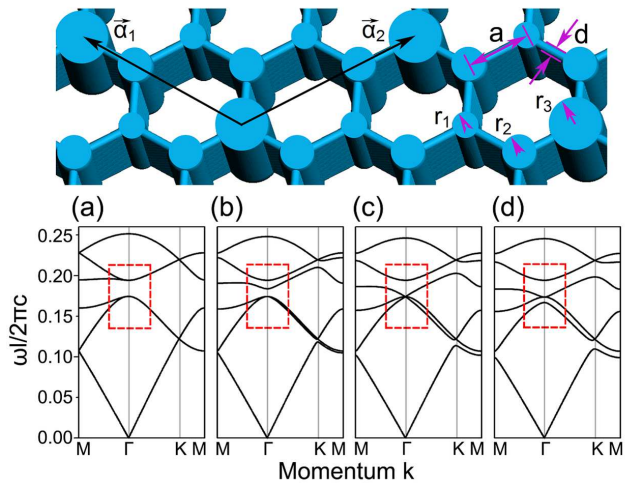}
 \caption{(color online) Upper panel: Schematic representation of honeycomb photonic crystals with a $\sqrt{3} \times \sqrt{3}$ periodicity. $\vec{\alpha}_1$ and $\vec{\alpha}_2$ denote the primitive vectors. The distance between nearest columns is set to be $a$ and the slab width is $d=0.1a$. $r_i$ ($i=1$-$3$) label the radii of  different columns. In our simulation, each column is chosen to be infinitely long. Lower panel: Photonic band structures of transverse-magnetic modes along high-symmetry lines for different radii $r_1=0.18a$ (a), $0.23a$ (b), $0.28a$ (c), and $0.32a$ (d), respectively. Here, we set $r_2=0.25a$ and $r_3=0.18 a$.
}\label{figure3}
\end{figure}

\textit{Photonic-crystal bands---.} Experimental realizations of periodic graphene superlattices could take advantage of substrates that can generate the $3\times 3$- or $\sqrt{3}\times \sqrt{3}$-type superstructure, like EuO(111)~\cite{HXYang} and Ag(111) substrates~\cite{Huang2012}. There are also other methods for engineering such kind of superstructures, \textit{e.g.}, silicene on Ag(111) substrate~\cite{Chen04}, InSb(111) surface~\cite{Nishizawa}, artificial organic molecular lattice~\cite{ZFWang}, or patterned two dimensional electron gas with well-established experimental technique~\cite{Park}. The applicability of our theory depends on the degree of the achievable commensurability with the crystal structure of honeycomb lattices. It is noteworthy that, since our model is spin independent, it can also apply to Bosonic systems like cold atoms~\cite{coldatom} or photonic crystals~\cite{pht2} in honeycomb superlattices. One possibility is to use honeycomb photonic crystals made of silicon columns linked by thin silicon slabs as shown in the upper panel of Fig.~\ref{figure3}, and use electromagnetic waves with transverse-magnetic modes in the $xy$ plane. The corresponding site potentials and hopping energies for the photonic crystal setup can be controlled through the column radius $r$ and the link width $d$. The confinement radii allow to tune the concentration of electrical-field energy of the harmonic modes.

If the columns' radii are identical and the connecting slabs have the same width, the two dimensional photonic band structure for transverse-magnetic modes~\cite{SeeSI} obtained from the finite elements method~\cite{pht0,pht1} shows two linearly dispersing Dirac cones, closely resembling the band structure of pristine graphene~\cite{SeeSI,pht2,pht3}. To model the $\sqrt{3}\times \sqrt{3}$ graphene supercell, we first classify the columns' radii
into three categories $r_i$ ($i$=1-3) with $r_3=0.18a$ as a reference, and the link width is chosen to be $d=0.10a$ with $a$ being the distance between two nearest columns. Figures~\ref{figure3}(a)-\ref{figure3}(d) display the photonic band structures for different $r_1=0.18a$ (a), $0.23a$ (b), $0.28a$ (c), and $0.32a$ (d) at fixed $r_2=0.25a$ along high symmetry lines. One can observe a topological phase transition from an insulator to a single-valley metallic phase when $r_1$ is progressively increased [See the highlighted regions] in a way closely similar to the behavior of the electronic band structure shown in Fig.~\ref{figure2}.

\textit{Discussions and Conclusions---.} We presented the theory for the inter-valley coupling mechanisms in $\sqrt{3} \times \sqrt{3}$ graphene supercells that act as in-plane pseudo-Zeeman fields or pseudospin-orbit coupling. Both contributions can be used to tailor valley pseudospins of honeycomb lattices. Especially, the Dresselhauls-type valley-orbit coupling makes it possible to control the valley polarization via electric means. These valley coupling mechanisms have important implications in valleytronics, where the coherent control of valley polarization is yet a grand challenge due to the missing counterpart mechanisms of spin-orbit couplings or magnetic fields for spintronics. Moreover, our theory also suggests strategies for engineering single-valley electronic structure in conventional Dirac materials with two inequivalent degenerate valleys by folding them together. The single-valley phase can be manipulated by combining inter-valley couplings and imbalanced sublattice potentials originated from the inversion-symmetry breaking. By increasing the strength of inter-valley coupling from zero, a topological phase transition can take place from the quantum valley-Hall phase to a chiral single-valley metallic phase with quadratic band crossover that resemble the electronic structure of a \textit{half} Bernal-stacked bilayer graphene. A concrete proposal for such a single-valley phase is presented in honeycomb photonic crystals. We verified that all these findings can also be realized in $3 \times 3$ honeycomb supercells.

\textit{Acknowledgements---.} We acknowledge financial support from 100 Talents Program of Chinese Academy of Sciences, NNSFC (11474265, 11104069 and 11034006), DOE (DE-FG03-02ER45958) and Welch Foundation (F-1255). J.J. thanks the support from National Research Foundation of Singapore under its Fellowship program (NRF-NRFF2012-01). Q.N. is also supported by NBRPC (2012CB921300 and 2013CB921900), and NNSFC (91121004) during his leave at Peking University. The supercomputing center of USTC is acknowledged for computing assistance.

\begin{appendices}
\section{Supplemental Materials}
\section{Introduction}

In the main text, we presented the theoretical formulation for engineering a single-valley phase by folding K and K$'$ valleys of graphene into the $\Gamma$ point via the top-site adsorption in a $\sqrt{3}\times \sqrt{3}$ supercell. In the following we supplement the information in the main text by providing more explicit details to facilitate understanding. In Section II, we provide a detailed analysis of the phase transition by exactly diagonalizing the tight-binding Hamiltonian, where an interesting partner switching of the energy bands occurs along with the phase transition from quantum valley Hall insulator to a metallic phase wit quadratic band crossover. In Section III, we present the details about the calculation of the photonic band structure and show the Dirac cones of honeycomb photonic crystals with uniform radii of slabs. In the main text, we have only focused on the top adsorption that breaks the inversion symmetry. When the atoms are adsorbed at hollow or bridge sites, the inversion symmetry is preserved. For completeness, in Section IV, we present the physical origins of inter-valley couplings for hollow- and bridge-site adsorption and the corresponding band structures.

\section{Top Adsorption in $\sqrt{3}\times \sqrt{3}$ Graphene Supercell}
\begin{figure}
   \includegraphics[width=9 cm,angle=0]{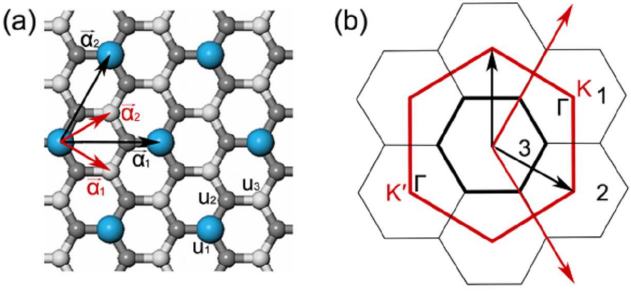}
   \caption{(color online) (a): Primitive cell for the $\sqrt{3}\times \sqrt{3}$ graphene supercell with top-site adsorption. $u_i$ ($i=1$-$3$) represent the site energies of the three type atoms. (b): Reciprocal lattices for $1\times1$ (in red) and $\sqrt{3}\times \sqrt{3}$ (in black) graphene supercells.}\label{top-root3}
\end{figure}

For the top adsorption, we choose $\sqrt{3} \times \sqrt{3}$ supercell as an example [See Fig.~S\ref{top-root3}(a)]. In this superlattice, if we only consider the site energy modification from the adatoms, all these site potentials $u_{i}$ can be classified into three types: $u_1$ for the sites right underneath the adatoms, $u_2$ for the sites nearest to adatoms, and $u_3$ for all other sites. For simplicity, $u_3=0$ is chosen as a reference. Therefore, the $\pi$-orbital tight-binding Hamiltonian can be written as:
\begin{eqnarray}
H_{\rm t} =    H_0 + u_1 {\sum_{i}}' a_{i}^{\dag} a_{i} + u_2 {\sum_{i}}'\sum_{\delta} b_{i+\delta}^{\dag} b_{i+\delta}, \label{eq1}
\end{eqnarray}
where ${\sum_{i}}'$ runs over all adatom sites. Here $H_0=- t_0 \sum_{<ij>}(a_i^{\dag}b_j + h.c.)$ is the band Hamiltonian with $t_0$ being the nearest-neighbor hopping energy, and $a_i^{\dag}$ ($b_i^{\dag}$) is the creation operator of an electron at $i$-th A(B) site. The second term is the site energy induced by adatoms possessing the translation symmetry of $\sqrt{3}\times \sqrt{3}$ supercell.

\begin{figure}
  \includegraphics[width=9 cm,angle=0]{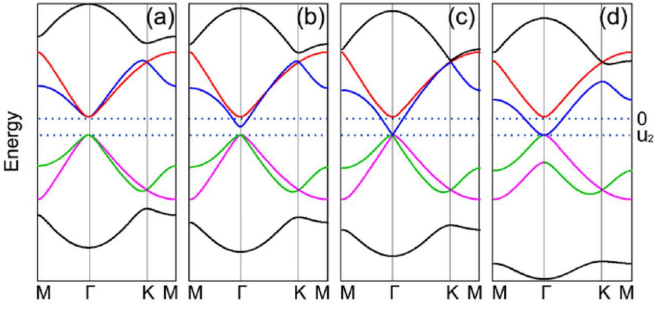}
  \caption{(color online) Evolution of tight-binding band structures for the $\sqrt{3} \times \sqrt{3}$ graphene supercell for different $u_1=0$ (a), $3u_2/4$ (b), $3u_2/2$ (c), $7u_2/2$ (d) at fixed $u_2<0$. Note: Bands' partner-switching happens in the last panel after the topological phase transition.} \label{full-band-root3}
\end{figure}

In the corresponding momentum space, the Brillouin zone of pristine graphene (red lines) is three times larger than that of the $\sqrt{3}\times \sqrt{3}$ supercell (black lines) are plotted in Fig.~S\ref{top-root3}(b). Thus, one can divide the Brillouin zone of pristine graphene into three parts with centers ${\bf K}_1$, ${\bf K}_2$, and ${\bf K}_3$ being wavevectors of K, K$'$, and $\Gamma$ points of graphene, respectively. Each part is the same as the Brillouin zone of $\sqrt{3}\times \sqrt{3}$ graphene supercell. Therefore, the operator $a_i$ can be expanded as:
\begin{eqnarray}    \label{fouriertranssqrt3}
    a_i = \frac{1}{\sqrt{N_0}} \sum_{\bm k} \sum_j \exp [{-i ({\bf{K}}_j+{\bm k}) \cdot {\bf{R}}_i }]a_{j,k},
\end{eqnarray}
where $\bm{k}$ runs over the Brillouin zone of $\sqrt{3}\times \sqrt{3}$ supercell. By substituting Eq.~\eqref{fouriertranssqrt3} into Eq.~\eqref{eq1}, the tight-binding Hamiltonian in momentum space can be expressed as follows:
\begin{eqnarray}\label{LowEnergyExpansion-k-s3xs3}
           H_{\rm{t}}(\bm{k}) =H_0(\bm{k})+ \sum_{j,j'} [\frac{u_1}{3} a_{j,k}^{\dag}a_{j',k} + \frac{u_2}{3} \xi_{jj'} b_{j',k}^{\dag}b_{j,k}].
\end{eqnarray}
where $H_0(\bm{k})=- t_0 \sum_{j}(\chi_{jk} a_{j,k}^{\dag}b_{j,k} + h.c.)$ is the kinetic energy of pristine graphene with $\chi_{jk}=\sum_{\delta}e^{-i({\bf{K}}_j+\bm{k})\cdot \bm{\delta}}$, and $\xi_{jj'}=\sum_{\delta}e^{i({\bf{K}}_j-{\bf{K}}_{j'})\cdot \bm{\delta}}$.
The last two terms give sublattice potentials when $j=j'$ which are different for AB sublattices due to inversion symmetry breaking. When $j\neq j'$, they furnish inter-valley coupling where $u_1$ contribution is finite while $u_2$ contribution vanishes due to the phase interference ($\xi_{KK'}$=0).

By exactly diagonalizing Eq.~\eqref{LowEnergyExpansion-k-s3xs3}, the tight-binding band structures for $\sqrt{3} \times \sqrt{3}$ supercells with a negative $u_2$ is plotted as shown in Fig.~S\ref{full-band-root3} for different $u_1$. One finds that when $u_1$ decreases from zero, the band gap from imbalanced sublattice potentials [See Fig.~S\ref{full-band-root3}(a)] gradually decreases due to the emergence of inter-valley coupling [See Fig.~S\ref{full-band-root3}(b)]. When $u_1$ further decreases to a critical point, the bulk band gap completely closes and a linearly dispersed Dirac-cone is formed centered at the edge of the parabolic dispersed valence band. For even smaller $u_1$, however, no bulk band gap reopens while a local band gap opens due to the increasing of inter-valley coupling. Such a local gap lifts the three-fold degenerate at $\Gamma$ point and form a chiral single-valley metallic phase with quadratic band crossover. Interestingly, along with the phase transition, the third and fourth energy bands represented by green and purple lines switch partners.

\begin{figure}
   \includegraphics[width=8cm,angle=0]{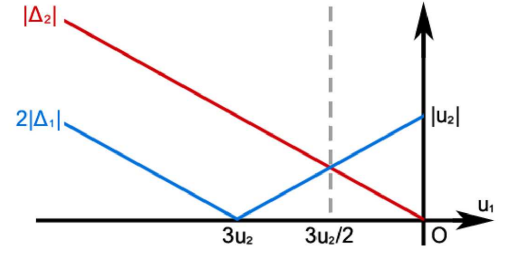}
   \caption{(color online) Schematic of the staggered sublattice potential $2|\Delta_1|$ and inter-valley scattering $|\Delta_2|$ as function of $u_1$.}\label{Fig4}
\end{figure}
To understand the phase transition process more clearly, we first obtain the effective Hamiltonian by utilizing L\"{o}dwin block diagonalization method. In this method, we first rearrange the Hamiltonian in Eq.~\eqref{LowEnergyExpansion-k-s3xs3} to divide it into two blocks: one $4\times 4$ matrix representing low energy bands contributed from valleys KK', and the other $2\times 2$ matrix representing high energy bands from $\Gamma$ valley of graphene. Their coupling is much smaller than the hopping energy $t_0$ and thus, the influence of high energy bands on the low energy ones can be obtained by perturbation method. Through a block diagonalization~\cite{Winkler}, the effective Hamiltonian can be obtained as below:
\begin{eqnarray}
           H^{\rm eff}_{\rm{t}}&=&U_0+ v_F(p_x \sigma_x + \tau_z p_y \sigma_y)+\Delta_1 \sigma_z \\ \nonumber
            &+& \frac{\Delta_2}{2} (1+\sigma_z) \tau_x, \label{eff}
\end{eqnarray}
where $U_0=(\Delta_2+u_2)/2$ with $\Delta_2=u_1/3$ and $\Delta_1=(\Delta_2-u_2)/2$. The third term reflects the imbalanced sublattice potentials, while the last term describes the inter-valley coupling occurring at sublattice ``A".

\begin{figure}
   \includegraphics[width=8 cm,angle=0]{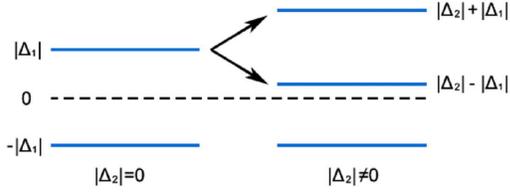}
   \caption{(color online) Left panel: Energy levels at $\Gamma$ point for $u_1=0$. Both energy levels are double degenerate. Right panel: Energy levels at $\Gamma$ point for $u_1=3u_2/4$. The degenerate of the upper energy level in left panel is lifted while the lower one remains degenerate.}\label{Fig3}
\end{figure}

To show the dependences of inter-valley coupling and imbalanced sublattice potentials on $u_1$, the magnitudes of $2\Delta_1$ and $\Delta_2$ are plotted in Fig.~S\ref{Fig4}. It shows that both terms depend on $u_1$ linearly. When the inter-valley coupling term is zero (\textit{i.e.}, $\Delta_2=0$), there are two double degenerate energy levels at $\Gamma$ point [See left panel of Fig.~S\ref{Fig3}]. The states at the upper (lower) energy level are contributed from ``A (B)" sublattices. When $u_1$ decreases from zero, the inter-valley coupling $|\Delta_2|$ increases from zero, while the sublattice potentials decrease as shown in Fig.~S\ref{Fig4}. As a result, the upper energy level split with a gap of $2\Delta_2$ due to the inter-valley scattering, whereas the lower one energy level is unchanged since the inter-valley scattering only occurs at sublattices ``A" as shown in the right panel of Fig.~S\ref{Fig3}. As the $u_1$ further decrease to the critical point, an intersection between $|2\Delta_1|$ and $|\Delta_2|$ indicates that the lower energy level splitting from sublattice ``A" reaches the two-fold degenerate energy levels of  sublattice ``B", for which the bulk band gap closes. When $u_1$ becomes even smaller, the inter-valley coupling is always dominant and the band gap cannot reopen again and a chiral single-valley metallic phase is formed.

\begin{figure}
  \includegraphics[width=8 cm,angle=0]{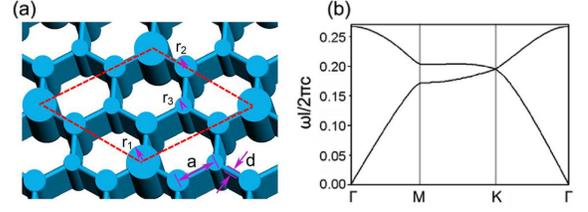}
  \caption{(color online) Left panel: Schematic of honeycomb-structured photonic crystals. The distance between nearest columns is set to be $a$ and the slab width is $d=0.1a$. $r_i$ ($i=1$-$3$) label the radii of  different columns. Right panel: The photonic band structure with uniform column radii and the slab widths where Dirac cones are present around K point and K$'$ point and the later one is not shown for clarity of the figure.} \label{lattice-photo}
\end{figure}

\section{Method for Calculating Band Structure of Photonic Crystal}
For the simulation of band structure in photonic crystals, we consider a honeycomb lattice with $\sqrt{3} \times \sqrt{3}$ periodicity, which is comprised of silicon columns linked with thin silicon slabs in the vacuum background as shown in the left panel of Fig.~S\ref{lattice-photo}. The columns are infinite in $z$-direction and their radii are $r_i$ ($i=1$-$3$). In our calculation, the electromagnetic wave propagates within the $xy$ plane, \textit{i.e.}, the wavevector component along $z$-direction is $k_z=0$. Various numerical methods can be used to calculate the photonic band structure, such as plane wave method (PWE), finite difference time domain (FDTD) method, and finite elements method (FEM) \cite{pht0,pht1}. Here, we used FEM, because it is much efficient in calculating structures with extremely small domains needing to be meshed. The photonic band structure with the uniform column radii and the slabs widths are calculated as shown in the right panel of Fig.~S\ref{lattice-photo}, where two Dirac cones in K and K' points for transverse-magnetic modes \cite{pht2,pht3} are formed, resembling the linear-dispersed Dirac cones of pristine graphene.

\section{hollow and bridge adsorption}

\begin{figure}
  \includegraphics[width=8 cm,angle=0]{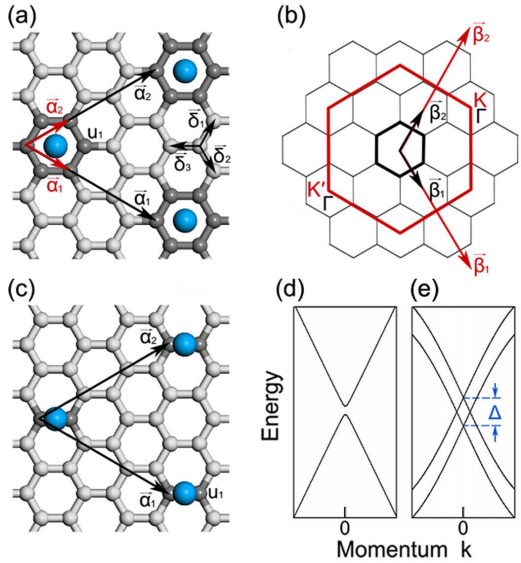}
  \caption{(color online) Primitive cells for hollow- (a) and bridge-site (c) adsorption. $\vec{\alpha_i}$ indicates the primitive lattice vectors for $1\times 1$ (in red) and $3\times 3$ (in black) graphene supercell. $u_1$ denotes the site-energy induced by the adatoms. (b): Reciprocal lattices for $1\times 1$ (in red) and $3\times 3$ (in black) graphene supercell. (d) and (e): Low energy band structures for hollow- and bridge-site adsorptions. $\Delta$ indicates the local gap from pseudo-Zeeman field in bridge adsorption. }\label{Figure1}
\end{figure}

In above, we study the top-site adsorption case where the inversion symmetry is broken and staggered AB sublattice potential is induced. In the following, we show the inter-valley coupling mechanisms for hollow and bridge adsorption cases with inversion symmetry in $3 \times 3$ honeycomb supercells. We first consider hollow adsorption as shown in Fig.~S\ref{Figure1}(a) with primitive vectors denoted by $\vec{\bm{\alpha}}_1$ and $\vec{\bm{\alpha}}_2$ in black. By considering only the site energies surrounding the adatoms, the $\pi$-orbital in real-space tight-binding Hamiltonian is written as:
\begin{eqnarray}    \label{3x3realtb}
           H_{\rm{h}} = H_0 + u_1{\sum_{i}}'(a_{i}^{\dag}a_{i} + b_{i}^{\dag}b_{i}),
\end{eqnarray}
where $\sum'_{i}$ runs over six atoms nearest to adatoms with site energy of $u_1$. Note that AB sublattices are equivalent here because of the inversion symmetry.

The reciprocal lattices for both $1\times 1$ (in red) and $3 \times 3$ (in black) graphene supercells are displayed in Fig.~S\ref{Figure1}(b). The Brillouin zone of pristine graphene can be divided into nine copies of that of $3\times 3$ graphene supercell. By denoting the $j$-th ($j=1$-$9$) center as ${\bf{K}}_j$, the operator $a_i$ can be expanded in momentum space as:
$a_i = \frac{1}{\sqrt{N_0}} \sum_{\bm k} \sum_j \exp [{-i ({\bf{K}}_j+{\bm k}) \cdot {\bf{R}}_i }]a_{j,k}$,
where $N_0$ is a normalization factor, and $\bm{k}$ runs over the Brillouin zone of $3\times 3$ graphene supercell. Therefore, the Hamiltonian of Eq.~\eqref{3x3realtb} in momentum space can be expressed as:
\begin{eqnarray}  \label{3x3momtb}
           H_{\rm{h}}(\bm{k}) = H_0(\bm{k})+\sum_{j,j'} \frac{u_1}{9} \xi_{jj'} (a_{j,k}^{\dag}a_{j',k} +  b_{j',k}^{\dag}b_{j,k}).
\end{eqnarray}
In the second term, $j=j'$ gives equivalent AB sublattice potentials, while $j \neq j'$ couples different parts. Although the direct coupling between valleys KK$'$ vanishes due to phase interference, $i.e.$, $\xi_{KK'}=\sum_{\delta}e^{i(\bf{K}-\bf{K}')\cdot \bm{\delta}}=0$, a band gap opens at $\Gamma$ point with four lower energy bands around the gap mainly contributed from eigenstates near valleys KK$'$. As shown in Fig.~S\ref{Figure1}(d). This suggests that the gap is induced by inter-valley coupling from higher-order effects. In below, we demonstrate the physical origin of inter-valley coupling for the hollow adsorption.

By doing a block diagonalization~\cite{Winkler} similar to the top-site adsorption case, a low-energy effective Hamiltonian at second-order approximation can be reached:
\begin{eqnarray}\label{LowEnergyExpansion-keff-3x3}
           H^{\rm{eff}}_{\rm{h}} = \frac{u_1}{3}+ v_F(k_x \sigma_x + \tau_z k_y \sigma_y) + \frac{u_1^2}{9t_0} \tau_x\sigma_y,
\end{eqnarray}
where $\bm{\tau}$ and $\bm{\sigma}$ are valley and sublattice Pauli matrices, respectively. The first term is an energy shift relative to the charge neutrality point, and the second term describes the kinetic energy with $v_F$ being the Fermi velocity. The last term couples valleys K and K$'$, where $\tau_x$ implies a pseudo-Zeeman field in $x$-direction to induce a procession of valley polarization. Moreover, the coupling only occurs between different sublattices, and the resulting band gap $2{u_1^2}/{9t_0}$ indicates a second-order correction from site energy $u_1$.

Then we study the bridge adsorption case as shown in Fig.~S\ref{Figure1}(c). Assuming that the adatom only influences the site energies $u_1$ of the nearest two carbon atoms and neglecting the high-order contribution from $\Gamma$ valley of graphene, the continuum effective Hamiltonian for four lower bands can be expressed as follows:
\begin{eqnarray}
           H^{\rm eff}_{\rm{b}} &=& \frac{u_1}{9}+ v_F(k_x \sigma_x + \tau_z k_y \sigma_y)+\frac{u_1}{9}\tau_x{\bm{1}}_{\sigma}. \label{bridge-eff}
\end{eqnarray}
The first and second terms are the energy shift and kinetic energy respectively, whereas the third term represents the first-order inter-valley coupling contributed from the on-site energy which also acts as a pseudo-magnetic field in $x$-direction yet without a sublattice flipping. This term shifts the two degenerate Dirac cones of graphene and opens a local energy gap $\Delta=2|u_1|/9$ at $\bm{k}=0$ as shown in Fig.~S\ref{Figure1}(e). The gap is closed at $(\pm |u_1|/9v_F,0)$ due to the dispersion of energy bands where another two Dirac cones are formed. All these results indicate that the inter-valley coupling mechanisms are sensitive to the geometry of adsorption.

\end{appendices}

\end{document}